\documentclass[aps,prl,twocolumn,groupedaddress,floatfix,showpacs,showkeys,amsmath,amssymb,10pt]{revtex4-1}
\usepackage{graphicx}
\usepackage{bm}
\begin{document}

\title{Semi-metallicity and electron-hole liquid in two-dimensional C and BN based compounds}

\author{Alejandro Lopez-Bezanilla$^{1}$}
\author{Peter B. Littlewood$^{2},^{3}$}

\email[]{alejandrolb@gmail.com}
\affiliation{Theoretical Division, Los Alamos National Laboratory, Los Alamos, New Mexico 87545, USA}
\affiliation{$^{2}$Argonne National Laboratory, 9700 S. Cass Avenue, Lemont, Illinois, 60439, United States}
\affiliation{$^{3}$James Franck Institute, University of Chicago, Chicago, Illinois 60637, United States}
\date{\today}

\begin{abstract}
Insulating-metallic transition mediated by substitutional atoms is predicted in a series of two-dimensional carbon-based structures. Introducing Si atoms in selected sites of tetrahexcarbon [Carbon 137 (2018) 266] according to rational chemical rules, metallicity by trivial band inversion without band gap opening is induced. Additional substitution of remaining C atoms by BN dimers introduces no changes in the metallic properties. 
A  series of isomorphous two-dimensional materials with isoelectronic structures derived by exchanging group IV elements exhibiting various band gaps is obtained. 
Dynamical stability is verified with phonon analysis and beyond the harmonic approximation with molecular dynamics up to room temperature. The semi-metallic compounds have well-nested pockets of carriers and are good candidates for the formation of an excitonic insulator.

\end{abstract}

\pacs{Valid PACS appear here}
\maketitle


\section{Introduction}
The successful synthesis of graphene\cite{Novoselov666} and numerous other two-dimensional (2D) materials marked the beginning of an acclaimed revolution in materials science. Much of the scientific and technological effort deployed on the field of two-dimensional (2D) materials is to ensure the creation of new compounds with outstanding electronic properties compatible with current synthesis techniques as in graphene.
However, chemical modification poses a design challenge in the design of new materials. For example, the conjugated $\pi$-network of graphene lattice allows it to exhibit metallic properties, relativistic dispersion of charge carriers, and ballistic transport\cite{Cresti2008}. These unique electronic features may be altered by modifying the graphene composition, even when its geometry and number of valence electrons are preserved. This is the case of massive graphene BN codoping\cite{BNcoDoping}, where polar bonds create electrostatic fields and potential variations at the atomic scale that disrupt graphene hyperconjugation. 

Richness or disorder in the composition may entail disruption of original properties \cite{Gregersen_2018} without necessarily obtaining new interesting properties.
As the intricacy of the novel compounds becomes more complex, a careful inspection of intermediate materials obtained in the process of discovery is revealed as a fundamental requirement to infer new material properties.    
Methodologies consisting in introducing modifications on existing structures guided by the intuition of accumulated experience are gaining relevance\cite{PIZZI2016218}. Indeed, examining the physico-chemical properties of the materials' constituents provides detailed insight of the interplay between elements in the arising of unexpected properties. Under this backdrop, accurate density functional theory (DFT) based calculations have enabled the prediction of multielement nanostructures with specific functionalities\cite{ISI:000427009000019}, facilitating the analysis of their properties before the actual synthesis.

In a previous paper\cite{pentaSiC2}, we presented a theoretical study of a new 2D material exhibiting a pentagonal arrangement of C and Si atoms that was obtained upon analysis of the atomic bonding of penta-graphene\cite{Zhang24022015}. In the buckled pentagonal structure of penta-graphene, some C atoms are fourfold coordinated with their atomic orbitals in sp$^3$ hybridization. Substituting those C atoms by Si atoms, the final nanostructure exhibits enhanced stability and new electronic properties. This is due to the ability of Si valence atomic orbitals to rehybridize in the sp$^3$ configuration, as the existence of silicene demonstrates\cite{PhysRevLett.108.155501}. A different hybridization is sp$^2$d, where interspersed fourfold coordinated Si atoms in graphene form stable structures \cite{PennycookSi, idrobo}.

\begin{figure}[htp]
 \centering
    \includegraphics[width=0.48\textwidth]{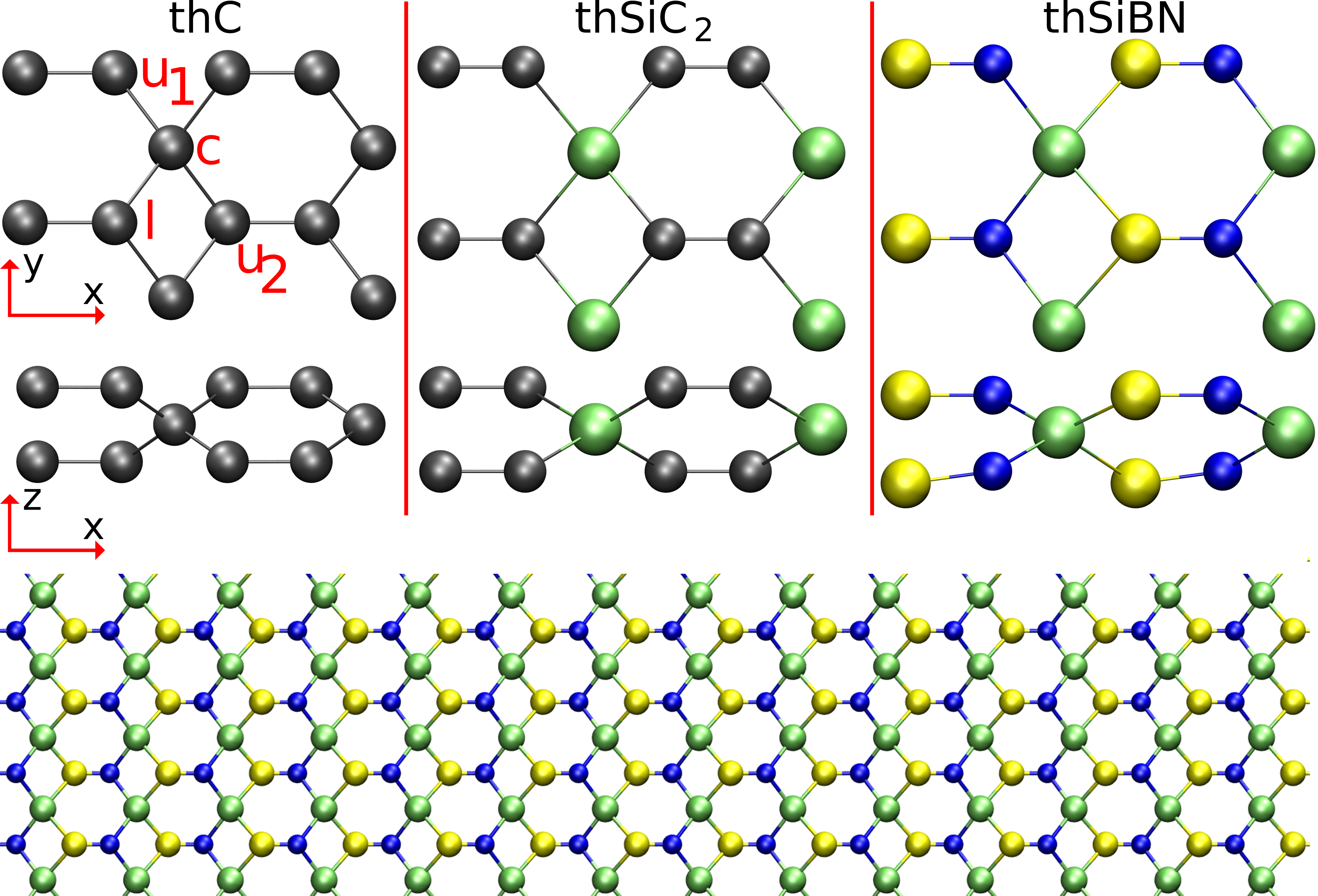}
 \caption{Top and side view of the schematic representation of the tetrahexC, tetrahexSiC$_2$, and tetrahexSiBN nanostructures. Angles between the central (c), lower (l) and two upper (u$_1$,u$_2$) atoms for each isomorph structure are given in Table \ref{table}. Lower panel displays the diamond and hexagon arrangement of tetrahexSiBN. Grey, green, yellow, and blue spheres represent C, Si, B, and N atoms respectively. }  
 \label{fig1}
\end{figure}

Following the same strategy, in this paper we report a series of buckled 2D compounds composed of four- and six-atoms rings obtained by chemical modification of tetrahexcarbon\cite{tetrahexcarbon,PhysRevMaterials.3.066002} (thC), a carbon allotrope obtained by applying a Stone-Wales transformation to penta-graphene\cite{pentagraphene}. Substitution of sp$^3$ hybridized C atoms by group IV atoms introduces a stress release in the nanostructure that expands the in-plane unit cell vectors, increases the buckling, and turns the insulating carbon allotrope into several forms of metallic and semiconducting materials. Additional substitution of C dimers by BN dimers yields tetrahexSiBN (thSiBN), a metallic 2D material.

\section{Geometric model of the nanostructures}

\begin{table}[htbp]
        \caption{Unit cell vectors (n$_x$, n$_y$) and angles between central (c), lower (l) and two upper (u$_1$,u$_2$) atoms (as shown in Figure \ref{fig1} of all the nanostructures. The maximum vertical separation between atoms is provided as the thickness of the buckled nanostructures.}
\begin{center}
\begin{tabular}{|r|c|c|c|c|c|c|}
\hline
\multicolumn{1}{|l|}{} & \multicolumn{1}{c|}{thC} & \multicolumn{1}{c|}{thSiC$_2$}& \multicolumn{1}{c|}{thGeC$_2$}& \multicolumn{1}{c|}{thSnC$_2$}& \multicolumn{1}{c|}{thCBN}& \multicolumn{1}{c|}{thSiBN} \\ \hline
n$_x$(\AA)                   &6.06 &  7.19  & 7.26 & 7.81  &6.24   & 7.32 \\ \hline
n$_y$(\AA)                   &4.50 &  5.59  &5.87  & 6.53  &4.48   & 5.47\\ \hline
u$_1$-c-u$_2$ ($^\circ$)    &135.3 & 138.6 &138.5 & 142.4 &134.9  & 137.3 \\ \hline
u$_1$-c-l($^\circ$)         &112.3 & 108.7 & 110.0& 108.6 &107.4  & 109.5\\ \hline
u$_2$-c-l($^\circ$)         &84.9  & 86.3  & 84.8 &  83.7 &  86.1 & 88.4\\ \hline
Thickness (\AA)                       & 1.19 &  1.36 & 1.39 & 1.41 & 1.27 & 2.61\\ \hline
\end{tabular}
\label{table}
\end{center}
\end{table}

Ram and Mizuseki designed monolayered thC \cite{tetrahexcarbon} by applying a periodic rotation of a C-C dimer (Stone-Wales transformation) to penta-graphene\cite{pentagraphene}. The resulting structure is composed of a periodic arrangement of buckled diamond and hexagons exhibiting dynamical and thermal stability. Two types of C atoms can be differentiated according to the hybridization of their orbitals, namely sp$^2$ and sp$^3$ atoms. Whereas the former are threefold coordinated and their hybridized orbitals form angles of $\sim$120$^\circ$, similar to graphene, the latter are fourfold coordinated and tend to adopt a tetrahedral geometry (similar to diamond) which is frustrated by the nearly planar configuration of the layer. This leads to a nanostructure that is higher in energy than that of graphene\cite{tetrahexcarbon}.

A natural way to alleviate the stress imposed by the frustrated tetrahedral geometry is to substitute the C sp$^3$-hybridized atoms by an atom  whose orbitals can either combine yielding a flat configuration or simply be more prone to hybridizing in the sp$^3$ configuration than a C atom. Si exhibits sp$^3$ hybridization in a 2D buckled structure (silicene) \cite{PhysRevLett.108.155501,verri1}, suggesting that a closer-to-tetrahedral arrangement of the Si atomic orbitals is preferred. Also, experimental observations of inclusions of Si atoms in graphene layers\cite{PennycookSi,idrobo} were reported and an sp$^2$d atomic hybridization proposed to explain the flat fourfold Si coordination. 
 
Figure \ref{fig1} shows the geometry of tetrahexSiC$_2$ (thSiC$_2$) where each Si atom is bonded at first neighbours to four C dimers. 
This buckled structure exhibits a formation energy 2.15 eV lower than that of the completely flat geometry, suggesting that the sp$^3$ atomic hybridization is preferred over the sp$^2$d. Although the shape of the unit cell and atomic arrangements are the same than in thC, cell vectors increase substantially (18.8\% in x and 24.4\% in y ) to accommodate the Si atoms, as shown in Table \ref{table}. Angles change accordingly as shown in Table \ref{table}. As discussed in section \ref{ep}, in addition to Si atom other larger size atoms of group IV are considered in substitution of sp$^3$ hybridized atoms, which yields larger unit cells

\section{\label{ep}Electronic properties}
\begin{figure*}[htp]
 \centering
   \includegraphics[width=0.99\textwidth]{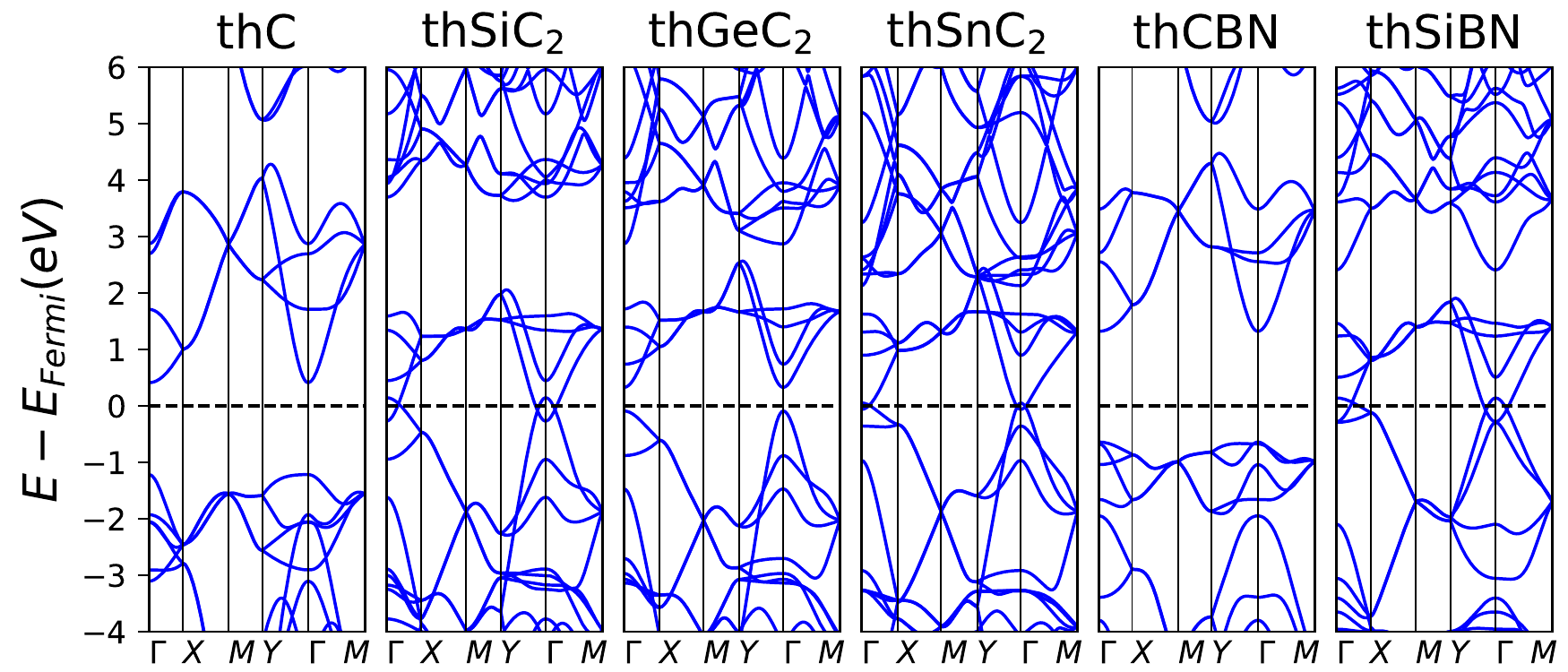}
 \caption{Electronic band diagrams of tetrahexC, tetrahexSiC$_2$, tetrahexGeC$_2$, tetrahexSnC$_2$, tetrahexCBN, and tetrahexSiBN. The insulating behavior of the carbon allotrope becomes metallic upon Si substitution of the sp$^3$ hybridized C atoms. Substitution of C atoms by BN dimers introduces little modification in the band structure.}  
 \label{figBands}
\end{figure*}

\begin{figure*}[htp]
 \centering
   \includegraphics[width=0.99\textwidth]{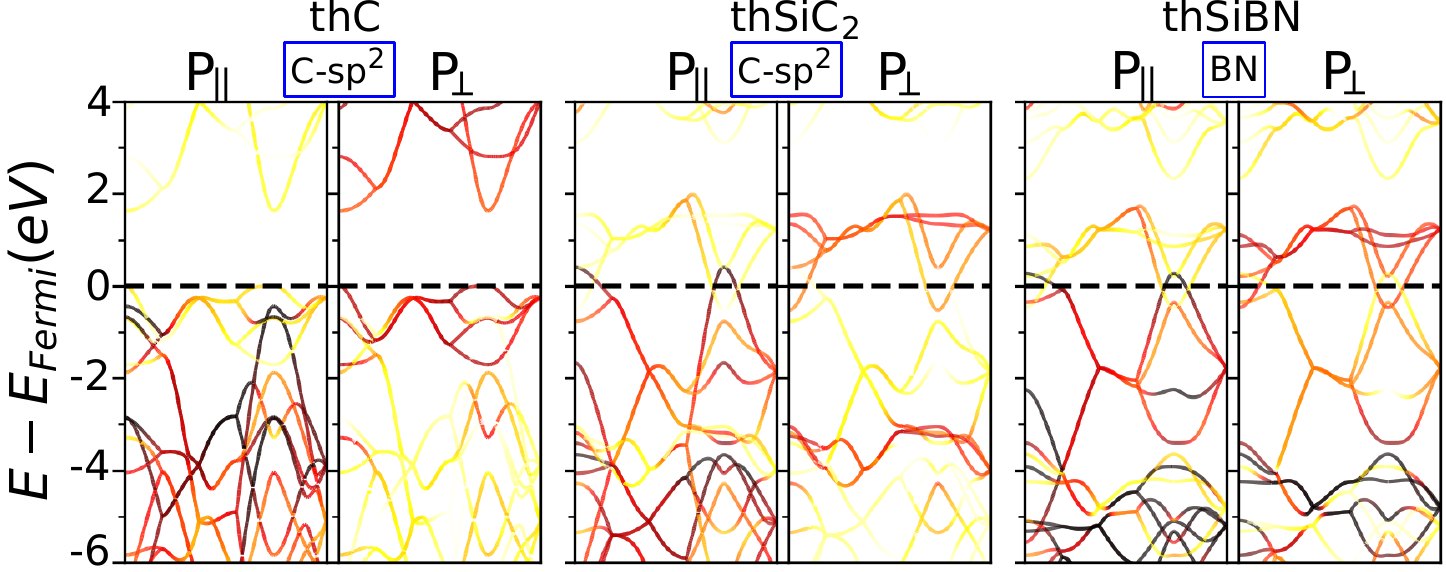}
 \caption{Color-weighted band diagrams of threefold coordinated atoms showing the  independent contribution of perpendicular to the plane p$_\perp$  and in-plane p$_{||}$  orbitals of tetrahexC, tetrahexSiC$_2$, and tetrahexSiBN. Fourfold coordinated atoms barely contribute to the formation of electronic states in the vicinity of the Fermi level. The color intensity of the lines is proportional to the contribution. }  
 \label{pdos}
\end{figure*}

The electronic properties of thC are dominated by the alternating arrangement of hexagons and diamonds that prevents thC from being hyperconjugated. Within the LDA approximation, the band diagram exhibits an electronic band gap at the Fermi level of 1.63 eV, as shown in Figure \ref{figBands}. The exact gap value may depend on the exchange-correlation functional used in the mean-field calculation\cite{tetrahexcarbon}. A detailed inspection of the contribution of each atomic orbital to the band diagram is conducted with color-weighted bands as plotted in Figure \ref{pdos}. Whereas the fourfold coordinated C atoms barely participate in the low-energy spectrum of thC, the C atoms in sp$^2$ hybridization are responsible for creating the electronic states in the vicinity of the Fermi level. Orbitals contribute differently depending on the their orientation: perpendicular to the plane p$_{\perp}$ orbitals contribute more than the in-plane orbitals (p$_{||}$) to both conduction and valence bands. 
  
Dimers of p$_{\perp}$ orbitals in one unit cell are inefficient in forming a chemical bond with neighbouring p$_{\perp}$ orbitals in the next cell due to the intervening  sp$^3$ hybridized C atoms; this results in low-dispersive states at the Fermi level and an insulating band gap. 
Note an isolated set of four empty conduction bands formed by the lateral overlap of p$_{\perp}$ orbitals of C sp$^2$ hybridized dimers. Separated from the rest of conduction bands by a 0.9 eV band gap, these bands exhibit a dispersion of $\sim$4 eV. As explained below, the shift in energy of these four bands as a result of chemical modification is the key factor that determines the metallic or insulating character of the modified nanostructures. 

\begin{figure*}[htp]
 \centering
   \includegraphics[width=0.98\textwidth]{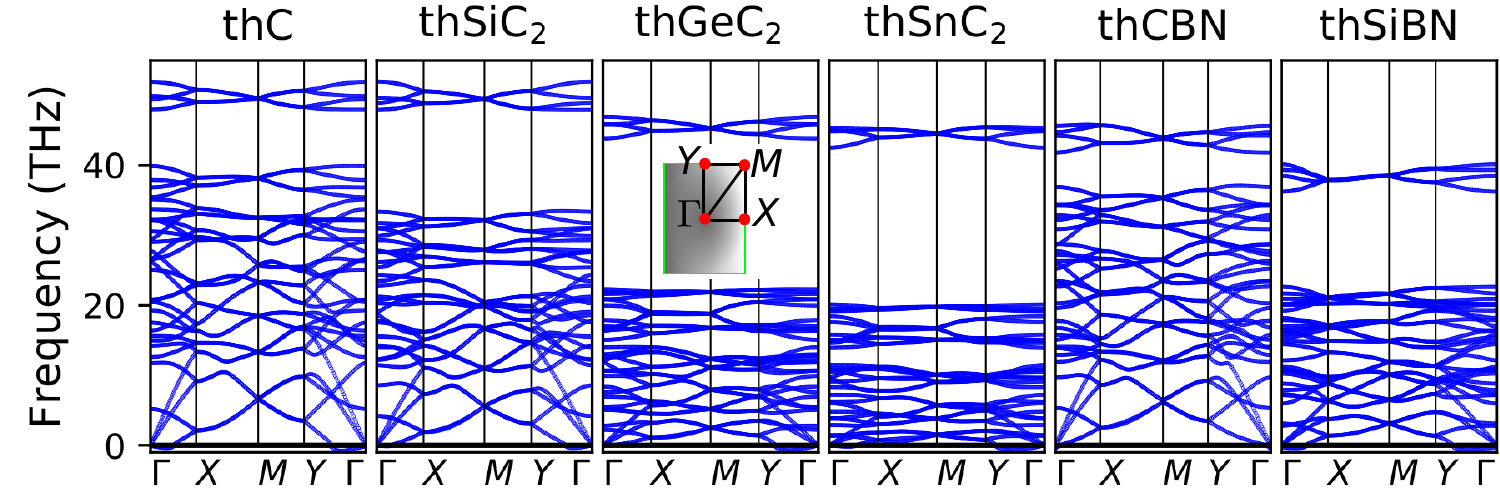}
 \caption{Phonon spectra of tetrahexC, tetrahexSiC$_2$, tetrahexGeC$_2$, tetrahexSnC$_2$, tetrahexCBN, and tetrahexSiBN. The frequency of each phonon mode is plotted versus its propagation direction along the high-symmetry lines in the Brillouin zone, as depicted in the inset.}  
 \label{figPhonons}
\end{figure*}

Si is one position below C in the group IV of the periodic table, namely both atoms share many characteristics such as the same number of valence electrons in a p-orbital: 2p in C atoms and 3p in the Si atoms. Going down in the group IV, heavy metal atoms such Ge and Sn also accommodate four electrons in their outermost 4p and 5p-orbitals respectively. As a result of the increasing  difference between the s and p valence orbitals as the size of the atoms increase, $\sigma$-orbitals and $\pi$-orbitals exhibit a non-negligible mixing in two-dimensional structures based on Si\cite{PhysRevLett.108.155501,2053-1583-3-1-012001}, Ge\cite{germanene14}, and Sn\cite{stanene15}. This leads to honeycombed structures where sp$^3$ hybridization is a configuration energetically more stable with respect to all-sp$^2$ graphene which favors the puckering of the layers. With the sp$^3$ hybridization being highly favorable, chemical substitution of sp$^3$ hybridized C atoms by Si, Ge, and Sn atoms is considered in the following as alternative materials. 

Figure \ref{fig1} shows the unit cell of tetrahexSiC$_2$ (thSiC$_2$) where the Si atoms are bonded to four C atoms. The same schematic representation is valid for Ge and Sn substituting atoms, although both the unit cell vectors and the bonding distances increase with increasing group IV atom size. The electronic band diagram of thSiC$_2$ in Figure \ref{figBands} shows that atomic substitution renders the nanostructure metallic, reduces the dispersion of the conduction bands, and leads to a trivial band inversion by which two of the valence bands become partially empty at the expense of filing partially two conduction bands. The insulating-metallic transition is driven by a shift down in energy of the conduction bands (with respect to thC), that increase their gap to $\sim$2 eV with the higher bands, and by a shift up in energy of a set of two occupied bands. 

An inspection of the color-weighted bands plotted in Figure \ref{pdos} shows that the cone-shape band that in thC was totally filled and formed by the in-plane p-orbitals shifts in energy and becomes partially empty in thSiC$_2$. The cone-shape conduction band formed almost exclusively by p$_\perp$-orbitals compensates and produces an electron pocket. Near the zone center two bands play a similar role in the metallicity of the this structure. A charge transfer from p$_{||}$ to p$_\perp$ orbitals occurs as a result of introducing the Si atom in the nanostructure. Calculations including spin-orbit coupling leave the non-relativistic results unaltered.  

Si substitution by Ge atoms leads to thGeC$_2$. A major change introduced in the electronic band diagram when compared to the one of thSiC$_2$ is the semiconducting band gap of 400 meV at the $\Gamma$ point. Both the shape and dispersion of the conduction bands remain the same. It is interesting to point out that substitution of Si by Ge with no relaxation of the structure leads to a band gap of only 14 meV. (Similarly, substitution of C by Si or Sn shifts conduction bands away from valence bands without opening a band gap). Therefore, the combined effect of atomic substitution and structural relaxation adjust the level of the electronic states and the subsequent creation or disappearance of a band gap. 

Substitution of sp$^3$ atoms of thC by Sn atoms yields thSnC$_2$. As in thSiC$_2$ the gap vanishes and a small band inversion at $\Gamma$ point renders the nanostructure metallic. The electronic gap above the set of four conduction bands also closes and no major variation of the electronic band dispersion is observed.  

The non-monotonic trend of gap closing and opening in the sequence C-Si-Ge-Sn is on the first sight surprising, because it runs counter to the usual trend of gap reduction and closing in tetrahedrally coordinated semiconductors as the s-p splitting increases on moving down the periodic table. That trend is in fact present in the manifold of $\sigma^*$ antibonding orbitals which lies above the almost isolated $sp_\perp ^2$. In the most simplified view the $sp_\perp^2$ band from three-fold coordinated atoms is sitting rather like an ``impurity band'' of dimers in a wider gap insulator.  The bonding-antibonding $\sigma$ $\sigma^*$ manifold  is constructed from hybridised four-fold $sp^3$ and three-fold $sp_{||} ^2$; that gap shrinks monotonously from 6 $eV$ in thC to around 2 $eV$ in thSnC$_2$. The $p_\perp$ band sits at a relatively fixed energy as the larger gap collapses around it. 
   
A common characteristic of the four structures presented above is the isoelectronic character of their constituents. The number of valence electron remain constant and hence the band structures retain a similarity across the substitution series. An additional isoelectronic substitution in thC could change the C atoms in sp$^2$ hybridization. Given that 2D BN is an isomorph of graphene\cite{Golberg} able to integrate the layered structure in codoping, the C dimers of the four nanostructures could in principle be substituted by BN dimers. It is worth remembering that BN-based structures are wide-gap insulators due to the polar and ionic character of the B-N bond. 2D hexagonal BN was observed to show metallic behavior when cut along zigzag nanoribbons, which was ascribed to half-metallicity derived from the edge states\cite{Golberg}.

In Figure \ref{figBands} the band diagram of tetrahexCBN (thCBN) shows two sets of isolated four bands at both sides of the Fermi level separated by a 1.95 eV wide band gap. 
The most appealing BN substitution occurs in the dimers of thSiC$_2$, which yields tetrahexSiBN (thSiBN). Figure \ref{figBands} shows that thSiBN is metallic with a compensated electron-hole pocket similar to that of thSiC$_2$. The dispersive band in the $Y-\Gamma$ region vanishes, although the cone-shape band remains yielding band inversion in the zone center. An analysis of the weighted bands in Figure \ref{pdos} shows that the combined p$_\perp$ orbitals of both B and N atoms in the dimers contribute to the conduction bands more than the p$_{||}$ orbitals, and vice versa for the valence bands. As in the previous structures, the weight of sp$^3$ hybridized atoms in the formation of electronic states in the vicinity of the Fermi level is negligible. 

A qualitative Mulliken charge analysis reveals a strong ionic character in multi-atom compounds, specially both thSiC$_2$ and thSiBN. Whereas Si atoms in the former accept $\sim$0.1 electrons from the C atoms, in the latter each Si atom behaves as a donor of $\sim$0.3 electrons to each BN dimer.  

\section{Dynamical stability}

The first report on thC\cite{tetrahexcarbon} demonstrated the stability of the monolayer based on a phonon spectrum analysis. The absence of imaginary frequencies indicates dynamical stability since only restoring forces act on an atom that is perturbed a short distance from its equilibrium position. Tiny deviations observed in all structures were minimized within the LDA approximation and very tight convergence criteria in the first-principles calculations. Phonon calculations displayed in Figure \ref{figPhonons} shows deviations of $\sim$0.75 THz towards imaginary values for one of the acoustic branches starting at $\Gamma$ point. Whereas a deviation in between two points or at a point of the zone edge could indicate a structural phase transition, phonon softening at the $\Gamma$ point is indicative of some numerical inaccuracy, and does not demonstrate a structural instability. All compounds presented are dynamically stable and restoring forces are efficient against small perturbation and maintain the structural integrity. 
Structures formed by substitution of C and Si atoms by Ge and Sn atoms in BN-based structures exhibit instabilities and were discarded in this study.

\begin{figure}[htp]
 \centering
   \includegraphics[width=0.48\textwidth]{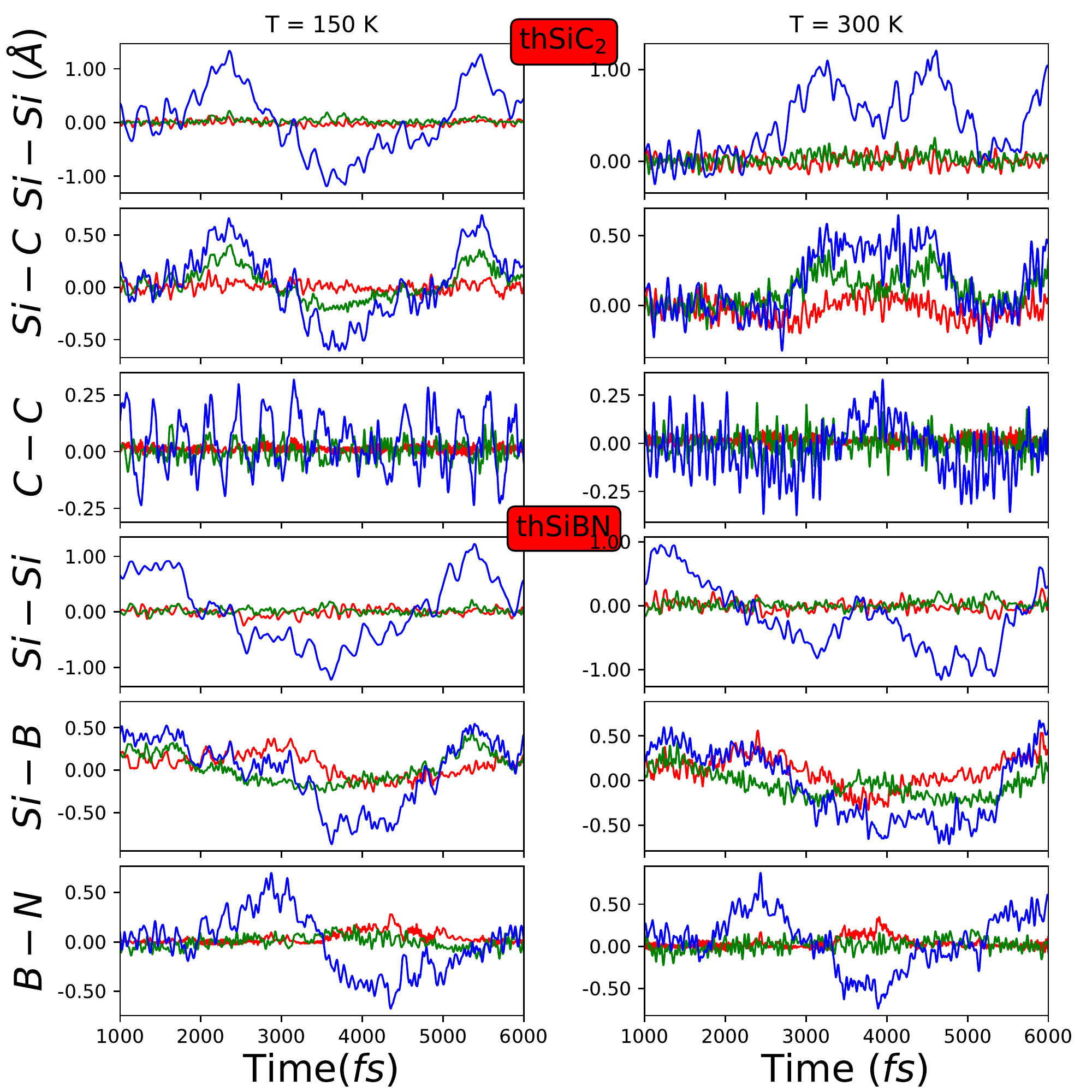}
 \caption{Molecular dynamic simulations at 150K and 300K demonstrate structural stability of thSiC$_2$ and thSiBN beyond the harmonic approximation at 0K. Variations of distances between atoms in the three perpendicular directions between pairs of atoms remain within a 1 \AA\ limit for the out-of-plane direction throughout 5 ps after an 
 time of 1 ps. Blue lines indicate out of plane displacement while red and green in-plane x and y displacement respectively.}  
 \label{figMD}
\end{figure} 

A total of 32 acoustic and optic modes extend over a frequency range of $\sim$40 THz. Separated by a $\sim$8 THz gap, four additional optical modes corresponding to in-plane three-fold coordinated atom vibration, lay at $\sim$50 THz. This gap increases when the difference of the atomic masses becomes more pronounced in thSiC$_2$ and thSiBN. Also the phonon dispersion lowers with the 32 branches spanning a range of $\sim$25 THz. 

The structural integrity of the layered structures is further verified with molecular dynamic calculations at finite temperature. Starting from the ground-state at T = 150 K and T = 300 K, 5 ps long runs were performed after an initial equilibration time of 1 ps with a 1 fs time-step. From heating at constant temperature the two-dimensional slabs composed of 192 atoms for 5 ps, no structural rearrangement of the atoms was observed. Figure \ref{figMD} shows relative interatomic distance variations in each orthogonal direction, of nearest Si-Si atoms, and Si-C and C-C dimers. The longer distance variation between atoms is the perpendicular-to-the-plane direction (up to 1 \AA), whereas atoms modify their relative distances in the structure plane in a much smaller range. The collective motion of the atoms is a breathing vibration of the nanostructure that expands and compresses, with no large variation of the bonding distance between pairs of atoms. Similar structural stability was observed for the rest of the compounds at finite temperatures. This demonstrates that the structural integrity of the nanostructures is preserved beyond the harmonic approximation, confirming that the compounds are stable up to room temperature. 

An alternative arrangements were considered corresponding to a BN dimer followed by a NB dimer along the x-axis after the C bond, and then the opposite arrangement from the parallel BN-C-NB line. That yields two N and two B atoms in front of each other at each side of a C atom, which reduces the distance between two atoms of the same kind, increasing the dipole and destabilizing the structure. The B-C-B and N-C-N distances would be 1.96 \AA and 2.06 \AA, as opposed to 2.1 \AA of B-C-N. In the structures studied here, the shortest distance between two B or two N  atoms is 2.5 \AA, which helps to balance the negative repulsion between electronegative N atoms. In addition, phonon calculations reveal important softening of some modes as a result of the atomic repulsion.

In a layered structure, the z-axis acoustic modes are relatively soft, and the z-axis lattice constant is potentially temperature-dependent and of course difficult to estimate from numerics with high accuracy. This raises the question about whether the semimetallicity might be dependent on the c-axis lattice constant. The chemical trends elucidated in section III around the closing of the gap between $\sigma$ and $\sigma^*$ orbitals is robust. The overlap of the bands is determined by the bandwidth of the dimer band relative to the bonding-antibonding gap. Both of these parameters are established by local in-plane atomic correlations which are not strongly affected by the interlayer distance in the simulation.

\section{Conclusions}
To summarize, a series of 2D materials composed of C and BN dimers woven in a hexagonal-square motif with group IV elements has been studied. Despite the isomorphism and isoelectronic features of the structures, a wide range of electronic behaviors are found. Exchanging C by BN dimers and substituting fourfold coordinated C atoms by isoelectronic Si, Ge, and Sn atoms, insulating-metallic transitions are observed. Whilst the carbon allotrope exhibits an electronic band gap, substitutional of sp$^3$ hybridized C atoms by Si or Sn atoms brings together conduction and valence bands leading to band inversion. Ge atoms yield a similar band structure to the one composed with the previous atoms but exhibit a small band gap.
It is remarkable that a 2D nanostructure composed of three different elements exhibits a metallic band structure, since the polar bonds typically lead to gaped electronic band diagrams.

Finally we remark that the electron-hole pockets that form at the fermi surface are very well nested, and the electron gas parameter $r_s = 1/(k_F a_B)$ is about 5 (here $k_F$ is the Fermi momentum and $a_B$ the Bohr radius. Another way to put this is that the Fermi energy of the electron (hole) pockets is around 0.1 eV, which is considerably smaller than the (unscreened) Coulomb energy $e^2/r_s$ of order 2 eV. Since the charge transfer from the in plane (hole pocket) to out of plane (electron pocket) will generate a Coulomb interaction that is relatively weakly screened, these semimetals will be good candidates for an excitonic insulator instability where the electron-hole pairs condense.\cite{zhu95}

\section{Computational Methods}
Self-consistent density functional theory (DFT) based calculations were performed within local density approximation (LDA) approach\cite{PhysRev.140.A1133} for the exchange-correlation functional was used as implemented in the SIESTA code\cite{0953-8984-14-11-302, PhysRevB.53.R10441,siesta2020}. A double-$\zeta$ polarized basis set was used to relax atomic coordinates and compute phonon spectra. 
Norm-conserving pseudopotentials as implemented in the SIESTA code were employed to represent the electron-ion interaction, so that the pseudopotentials account for the effect of the Coulomb potential created by the nuclei and core electrons creating an effective ionic potential acting on the valence electrons\cite{RIVERO201521}.
Atomic positions of the layered materials were fully relaxed with a force tolerance of 0.01 eV/\AA. The integration over the Brillouin zone (BZ) was performed using a Monkhorst sampling of 38 $\times$ 42 $\times$ 1 k-points. The radial extension of the orbitals had a finite range with a kinetic energy cutoff of 50 meV. A vertical separation of 35 \AA\ in the simulation box prevents virtual periodic parallel layers from interacting. The force-constant method and the PHONOPY package \cite{phonopy} were employed for computing phonon spectra. DFT-based molecular dynamic simulations with a Nos\'e thermostat\cite{doi:10.1063/1.447334} were performed in the NVT ensemble for a 4$\times$4$\times$1 unit supercells using only the $\Gamma$ point.

\section{Acknowledgments}
Los Alamos National Laboratory is managed by Triad National Security, LLC, for the National Nuclear Security Administration of the U.S. Department of Energy under Contract No. 89233218CNA000001. 
We acknowledge the computing resources provided on Bebop, the high-performance computing cluster operated by the Laboratory Computing Resource Center at Argonne National Laboratory. 
Work at Argonne is supported by Department of Energy, Office of Science, Basic Energy Sciences Division of Materials Science under Contract No. DE-AC02-06CH11357.

\end{document}